\documentclass{vgtc}                         % final (conference style)
\ifpdf%                                % if we use pdflatex
  \pdfoutput=1\relax                   % create PDFs from pdfLaTeX
  \usepackage{graphicx}                % allow us to embed graphics files
  \DeclareGraphicsExtensions{.pdf,.png,.jpg,.jpeg} % for pdflatex we expect .pdf, .png, or .jpg files
\else%                                 % else we use pure latex
  \usepackage{graphicx}                % allow us to embed graphics files
  \DeclareGraphicsExtensions{.eps}     % for pure latex we expect eps files
\fi%

\graphicspath{{figures/}{pictures/}{images/}{./}} % where to search for the images

\usepackage{microtype}                 % use micro-typography (slightly more compact, better to read)
\PassOptionsToPackage{warn}{textcomp}  % to address font issues with \textrightarrow
\usepackage{textcomp}                  % use better special symbols
\usepackage{mathptmx}                  % use matching math font
\usepackage{times}                     % we use Times as the main font
\usepackage{cite}                      % needed to automatically sort the references
\usepackage{tabu}                      % only used for the table example
\usepackage{booktabs}                  % only used for the table example
\usepackage{enumitem}
\usepackage{tabularx}
\usepackage[T1]{fontenc}
\usepackage[font=small,labelfont=bf]{caption}
\usepackage{todonotes}
\usepackage{multirow}
\usepackage{amsmath}

\onlineid{1207}

\vgtccategory{Research}

\vgtcinsertpkg

\title{Light Virtual Reality Systems for the Training of Conditionally Automated Vehicle Drivers}

\author{Daniele Sportillo \thanks{e-mail: daniele.sportillo@\{mpsa.com,mines-paristech.fr\}}
\and Alexis Paljic\thanks{e-mail: alexis.paljic@mines-paristech.fr}
\and Philippe Fuchs\thanks{e-mail: philippe.fuchs@mines-paristech.fr}}
\author{Daniele Sportillo \thanks{e-mail: daniele.sportillo@\{mpsa.com,mines-paristech.fr\}}\\ %
       \parbox{1.15in}{\scriptsize \centering PSA Group\\ MINES ParisTech, PSL\\ Centre for robotics}
\and Alexis Paljic\thanks{e-mail: alexis.paljic@mines-paristech.fr}\\ %
       \parbox{1.15in}{\scriptsize \centering MINES ParisTech,  PSL\\Centre for robotics}
\and Luciano Ojeda\thanks{e-mail: luciano.ojeda@mpsa.com}\\ %
      \parbox{1in}{\scriptsize \centering\scriptsize PSA Group}
\and Philippe Fuchs\thanks{e-mail: philippe.fuchs@mines-paristech.fr}\\ %
       \parbox{1.15in}{\scriptsize \centering MINES ParisTech,  PSL \\ Centre for robotics}
\and Vincent Roussarie\thanks{e-mail: vincent.roussarie@mpsa.com \newline2018 IEEE Virtual Reality (VR) \copyright 2018 IEEE\vspace{-0.5cm}}  \\ %
     \parbox{1in}{\scriptsize \centering PSA Group}}

\abstract{In conditionally automated vehicles, drivers can engage in secondary activities while traveling to their destination. However, drivers are required to appropriately respond, in a limited amount of time, to a take-over request when the system reaches its functional boundaries. In this context, Virtual Reality systems represent a promising training and learning tool to properly familiarize drivers with the automated vehicle and allow them to interact with the novel equipment involved. In this study, the effectiveness of an Head-Mounted display (HMD)-based training program for acquiring interaction skills in automated cars was compared to a user manual and a fixed-base simulator. Results show that the training system affects the take-over performances evaluated in a test drive in a high-end driving simulator. Moreover, self-reported measures indicate that the HMD-based training is preferred with respect to the other systems.}

\begin{document}

\firstsection{Introduction}

\maketitle

Conditionally automated vehicles do not require drivers to constantly monitor their driving environment; drivers can, therefore, engage in secondary activities such as reading, writing emails and watching videos. However, when the automated system encounters unexpected situations, it will assume that drivers will adequately respond to a Take-Over Request (TOR).

In this context, Virtual Reality (VR) constitutes a potentially valuable learning and skill assessment tool because it would allow drivers to familiarize themselves with the automated vehicle and interact with the novel equipment involved in a free-risk environment. 

The objective of this research is to explore the potential of \textit{light} Virtual Reality systems, in particular, their role for the acquisition of skills for the Transfer of Control (ToC) in highly automated cars. By using the adjective \textit{light}, we want to mark the difference between VR systems that are portable and/or easy to set up and systems that are cumbersome and require dedicated space to operate. The idea is that thanks to the portability and the cost-effectiveness, \textit{light} VR systems could be easily deployed in car dealerships to train a large amount of people in a safe and reliable environment. This paper aims to compare the effectiveness of a training program based on a head-mounted display with a user manual and with a fixed-base driving simulator. To validate the light VR system, user performances are evaluated during a test drive in a high-end driving simulator and self-reported measures are collected via questionnaires.

\subsection{Related Work}
%% VR DS %%
Although most of the studies in Driving Simulation (DS) uses static screens as the display system, recent studies prove that HMD-based DS leads to similar physiological response and lane change performance when compared to stereoscopic 3D or 2D screens \cite{weidner2017comparing} \cite{taheri2017development}.
Even if the steering wheel is the most used driving interface, novel HMD systems usually come with wireless 6-DoF controllers which can be used to control a virtual car \cite{sportillo2017immersive}. DS, in fact, provides the opportunity to implement critical scenarios and hazardous situations which are ethically not possible to evaluate on real roads \cite{ihemedu2017virtual}. For this reason and to overcome the limited availability of physical prototypes for research purposes, DS is extensively used for studies on automated vehicles to design future automotive Human-Machine Interface (HMI) \cite{melcher2015take} for TORs and to investigate the behavioral responses during the transition from automated to manual control \cite{merat2014transition}. Assessing the quality of the take-over performance remains an open problem. Usually reaction times (such as gaze reaction time, hands on wheel time, and intervention time), time to collision, lateral accelerations and minimum clearance are objective metrics analyzed in obstacle avoidance scenarios \cite{happee2017take}.

The intense scientific production in recent years suggests that the transition of control in automated cars is a valuable research topic worth investigating from the design stage to the final implementation of the new systems. Moreover, the compelling need and interest of the car industry to train a large amount of people in a reliable and cost-effective way, without compromising security, make light virtual reality system tools a promising solution for this purpose.

\section{Methods}
This study contained two parts: training and test drive. The aim of the training was to introduce the principles of the Level 3 Automated Driving System (ADS)-equipped vehicle, present the novel human-machine interfaces, and describe the actions to perform in order to appropriately respond to unplanned requests to intervene. A between-subject study with 60 participants was designed in order to compare a light Virtual Reality systems to a user manual and a fixed-base driving simulator in terms of training effectiveness evaluated through a test drive which required the application of knowledge and skills acquired during the training.

\subsection{The Training}
The aim of the training was to teach drivers how to interact with automated cars in three situations: the manual mode, automated mode and the take-over request. To do so, the training introduced the participants to the HMI for each situation, the actions they were free to perform during the automated driving and the best practice to respond to a take-over request. For all the participants, the training program started with an introductory video that briefly presented the main functionalities of a Level 3 ADS-equipped car. 

In the study three different training systems were compared. The first one was a user manual (UM) consisting of a slide presentation displayed on a 13.3" screen of a laptop computer. The participants were asked to carefully read each of the 8 slides and to proceed when they felt ready, without a time limit. The slides used text and images to present the actions to be performed during the manual driving, the automated driving and the take-over requests.

The second system was a fixed-base simulator (FB) consisting of an actual car cockpit including a driving seat, a dashboard, a force-feedback steering wheel and a set of pedals. A 9.7" tablet used by the driver to perform the secondary activity was placed in the center console. To display the virtual environment a 65" plasma screen was positioned behind the cockpit at 1.5m from the driver. 

The third system was a Light Virtual Reality (LVR) system including an HMD as a display system, and a racing wheel as driving system. Spatial sound was presented via headphones. To have a spatial correspondence between the real steering wheel and the virtual one, the steering wheel inside the virtual car was a 3D model of the real racing wheel with which the participants were interacting. Moreover, the position and the movements of the virtual model corresponded to the real one, allowing for co-located manipulation.

\subsubsection{The Virtual Learning Environment}
\begin{figure}[tb]
\centering
\includegraphics[width=\linewidth,trim={0 3cm 0 2cm},clip]{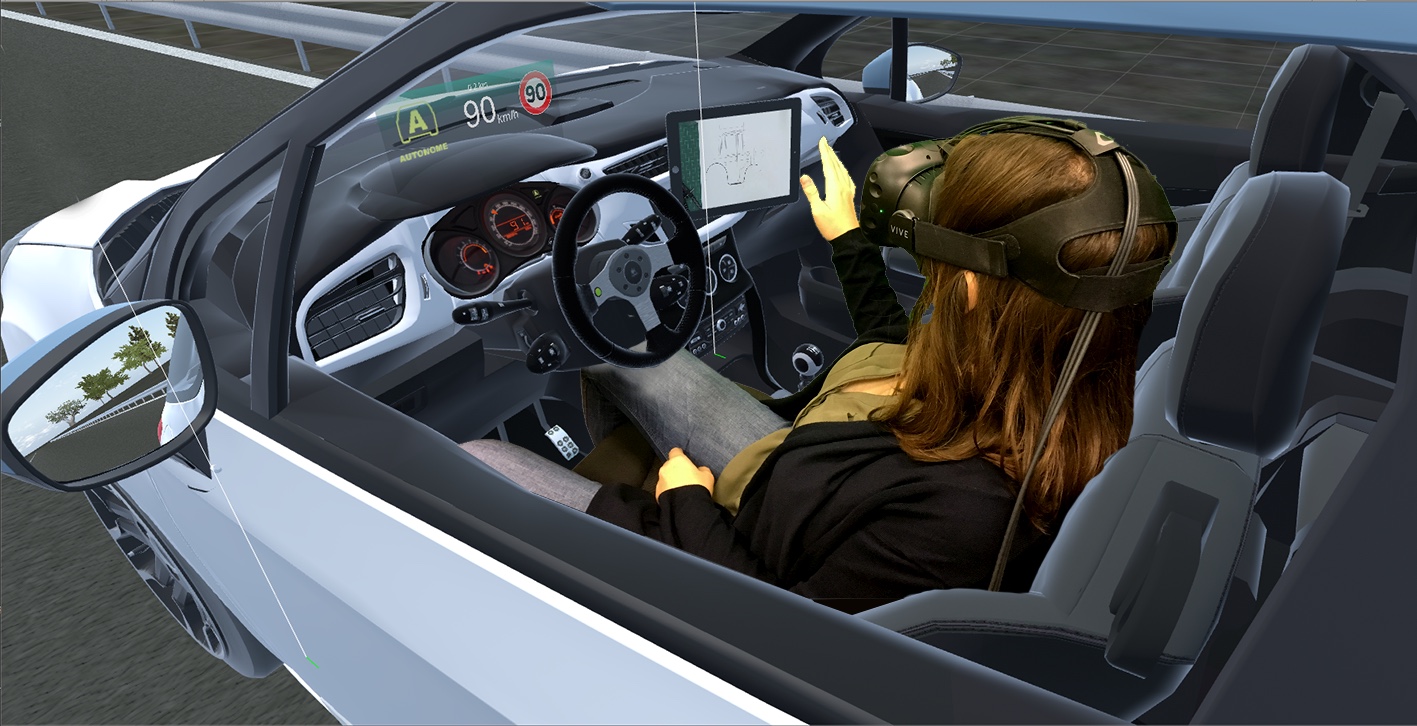}
\caption{Illustration of a user with HMD immersed in the Virtual Leaning Environment}
\label{fig:vle}
\end{figure}

For the training using the LVR system and the fixed-base driving simulator, a step-by-step tutorial was developed in the form of a Virtual Learning Environment (VLE) (\autoref{fig:vle}). The task of the participants consisted of interactions with the car following the instruction of a virtual vocal assistant. The messages announced by the assistant were also displayed on a yellow panel in front of the trainee. The driving scenario was a straight 2-lane road delimited by guardrails with no traffic. Before the driving scenario, an acclimatization virtual environment was proposed to the participants to help them locate and identify the controls of the car. This training also included a secondary activity (a video) that required the use of a tablet to distract the human driver from the driving task during the automated driving. The participants were asked, but not forced, to look at the tablet.
The video was automatically played when the automated system was enabled and paused during the manual driving and the TORs.

\subsection{Test Drive}
After the training, the participant performed a test drive designed to evaluate their performance in a more realistic driving scenario. The system used for this purpose was a high-end driving simulator consisting of the front part of a real car surrounded by a panoramic display. Data including position, speed and acceleration of the car, and current driving mode were recorded. Inside the car, a 10.8 inch tablet which provided 9 different secondary activities was placed in the center console. During the test drive 3 TORs were issued during the test drive: (A) a 10-second TOR caused by a road narrowing provoked by a stationary car on the right lane; (B) a 10-second TOR caused by a loss of ground marking; (C) a 5-second TOR caused by a sensor failure. During the autonomous driving, participants were asked to engage in one of the secondary activities proposed by the tablet.

\section{Results}
To evaluate the training systems and the learning environment, objective and self-reported measures were collected anonymously and treated confidentially. A first outcome of the study is that the training allowed all the participants to respond to the TORs.

The quality of the take-over was evaluated in terms of reaction time ($rt$), the elapsed time from the TOR until the driver takes back control. A significant difference between the user manual group ($\overline{rt}$ = 5.15s) and the other two systems was observed. However, no differences were observed comparing the fixed-base simulator ($\overline{rt}$ = 3.17s) and the HMD ($\overline{rt}$ = 3.16s). %This result suggests that participants who actually performed a take-over during the training were able to respond better to the first request to intervene in a more realistic driving situation. 
%It is important to point out that the reaction time of the second TOR of the participant trained with the user manual was significantly greater than the reaction time of the first TOR of the two other groups. 
Self-reported measures were collected through set of questions at the beginning of the test, after the training and after the test drive. To evaluate the favorability of the training, the participants filled out a 10 questions survey containing questions about perceived usefulness, easiness, pleasantness and realism. 
%Up to a total of 50 points (the highest the better), results showed that the HMD scored significantly better (M =  43) than both the fixed-base simulator (M = 40) and the user manual (M = 39.5). A Simulator Sickness Questionnaire (SSQ) was filled out by the participants who performed the training with the LVR system (N = 14.31, O = 7.58, D = 6.96).
In summary, according to the objective metrics measured during the test drive, the group of participants trained with the Virtual Learning Environment (fixed-base driving simulator and light VR system) were able to respond to the take-over request in less time than the group of participants trained with the user manual. Furthermore, self-reported measures showed responses in favor of the light VR training system. 

\section{Conclusion}
The results of this research persuade us that Light Virtual Reality systems represent a valuable tool for the acquisition of driving skills in conditionally automated vehicles. The proposed training system, composed of an HMD and a game racing wheel, is a portable and cost-effective device that provides an adequate level of immersion for teaching drivers how to respond to a take-over request in a safe environment. Therefore, this system could be employed for the training of future customers of automated cars before their first ride. A direct outcome of these results is the acknowledgment of VR as key player in the definition of the set of metrics for profiling driver interaction in automated vehicles. In future work, the training will be implemented in the form of a serious game in which the level of instruction adapts to the users' needs in order to assess the acquisition of skill during the training itself. 
%Additional work is necessary as well to further smooth the abrupt transition from the real to the virtual world caused by the HMD. 
Furthermore, test drives with the real vehicles are considered of primary importance to validate current results.

\acknowledgments{This research was supported by the French Foundation of
Technological Research under grant CIFRE 2015/1392 for the doctoral work of D. Sportillo at PSA Group.}
\bibliographystyle{abbrv-doi}

\bibliography{template}
\end{document}